\documentclass[twocolumn,prl,aps,showpacs]{revtex4}
\usepackage{graphicx}
\usepackage{dcolumn}
\usepackage{bm}
\usepackage{amsmath}
\begin{document}

\title{Theoretical unification between Quenched-Annealed and Equilibrated-Mixture Systems}
\author{R. Ju\'{a}rez-Maldonado$^{1}$ and M. A. Ch\'{a}vez-Rojo$^{2}$}
\affiliation{$^{1}$Unidad Acad\'emica de F\'{\i}sica, Universidad Aut\'onoma de Zacatecas. 
Calzada Solidaridad esquina con Paseo la Bufa S/N; Zacatecas, Zac. M\'exico.\\
$^{2}$Facultad de Ciencias Qu\'{\i}micas, Universidad Aut\'onoma de Chihuahua. 
Circuito No. 1, Nuevo Campus Universitario; Chihuahua, Chih. M\'exico.}
\date{\today}

\begin{abstract}
In this paper we apply the self-consistent generalized Langevin equation theory (SCGLE) 
of dynamic arrest for colloidal mixtures to predict the glass transition of a colloidal 
fluid permeating a porous matrix of obstacles with random distribution. We obtained the 
transition diagrams for different size asymmetries and so we give an asserted description 
of recent simulations results [K. Kim, K. Miyazaki, and S. Saito, Europhys. Lett. 88, 36002 (2009)]
of Quenched-Annealed and Equilibrated-Mixture systems which 
reveal very different qualitative scenarios which are in apparent contradiction with 
theoretical predictions of Mode Coupling Theory (MCT) [V. Krakoviack. Phys. Rev. E 75, 031503 (2007)]. 
We show that SCGLE theory predicts 
the existence of a reentrant region in EM systems as predicted using MC theory. However, 
opposite to MCT predictions, we show that it is practically impossible to distinguish a 
rentrant region in QA systems if it would exist. Qualitative comparisons are in good 
agreement with simulation results and thus, we propose SCGLE theory as a useful tool 
for the interpretation of the arrest transition in ideal porous systems.

\end{abstract}
\maketitle
The study of static and dynamic properties of fluids permeating porous materials is an interesting 
topic that has attracted increasing attention due to both its relevance in the study of 
systems in biology, chemistry, physics and engineering \cite{sahimi} 
and its scientific importance in the understanding of diffusing phenomena \cite{evans,confit}. 
For this, the 
development of theoretical schemes describing this class of systems has deserved considerable 
work from many perspectives. 
In this sense, a binary mixture in which one of the species remains immobile while the second 
species diffuse constitutes a model system that has been widely employed not only
in theoretical or simulation studies \cite{madden,gonzalo,porous,krakoviack,kim,kurzidim} 
but also in experimental works \cite{gil,kluijtmans} as an idealization of a 
fluid immerse in a porous medium.
Despite the simplicity of this model, some parameters 
like size asymmetry or matrix structure can be selected in order to mimic different systems. 
There are two protocols widely used to generate the obstacle matrix in order 
to simulate different classes of disordered porous materials,
known as Quenched-Annealed (QA) Mixture and 
Equilibrated Mixture (EM) \cite{kim}. 
In the first case, the porous matrix is prepared by quenching spherical particles of a monocomponent system 
in a typical equilibrium configuration and after that, mobile particles are inserted 
into the void spaces of the matrix of immobilized particles. By the opposite, in the EM protocol, 
all particles equilibrate together and then, a fraction of them are suddenly immobilized and so,
the structure of the obstacle matrix corresponds to that of one species of a 
mixture in equilibrium. 

On the other side, the study of the structural glass transition of fluids is another topic 
of interest because of its importance in the comprehension of dynamic phenomena as well as 
for its potential applications in industry. 
Nevertheless it has attracted a lot of interest in the last decades \cite{angell,stillinger,gotze}, 
there are still many open problems regarding to the nature of this dynamic phenomenon. 
For this, it is always desirable to have theoretical schemes that allow for the 
interpretation of experimental and computer simulation results. 
In this sense, Mode Coupling Theory (MCT) \cite{gotze,nagdhont,banchio,nagberdhont}
has shown its capability in predicting liquid-glass transitions 
in a variety of systems and conditions \cite{pham,szamel,tartaglia}. 
In this context, theoretical studies of liquid-glass transition of fluids confined in 
porous media deserves special interest in order to achieve a better understanding of 
the slow dynamics of these systems. 
Recently, based on the replica Orstein-Zernike 
equations \cite{madden,madden2,given}, Vincent Krakoviack has extended the 
Mode Coupling Theory to consider the glass transition of a fluid immersed 
in a porous medium \cite{krakoviack}. 
Within this Replica Mode Coupling Theory (RMCT) 
he obtained the dynamic arrest diagrams for QA mixtures and showed that
RMCT predicts a reentry phenomenon for matrix densities higher than the 
localization threshold. 
This contribution represents a crucial first step in the 
theoretical description of the slow dynamics of a fluid immersed in a porous medium.

Because of the complexity of the problem and the variety of scenarios regarding the 
glass transition in porous media, computer simulation studies have served to reveal 
some special features that allow to evaluate theoretical predictions. In particular, 
recent work of K. Kim and coworkers \cite{kim} has shown the strong dependence 
of the glass transition diagram on the protocol in which the porous matrix is prepared. 
In their work, they performed simulations of hard sphere systems permeating an obstacle
matrix prepared using both QA and EM protocols, and compared their results with 
RMCT predictions. Two important aspects should be addressed: a) by construction, 
there is no way of predicting EM glass transition diagrams using RMCT and
b) simulation results are in apparent contradiction with RMCT prediction 
about the reentrant behaviour in QA systems.

In this sense, the prediction of dynamic arrest transition diagrams employing 
an alternative theoretical scheme is crucial for the interpretation of the 
simulation results mentioned above. 
The Self Consistent Generalized Langevin Equation theory (SCGLE) has been developed 
in the context of colloidal systems \cite{laura1,laura2,simplified} and it has shown their capability in 
describing the dynamic properties of a variety of systems \cite{laura3}. 
The extension of this theory to mixtures has been carried out \cite{marco1,marco2} 
and allows us to consider QA an EM systems, just assuming that one species does not diffuse.
In fact, it has already been employed to describe the time evolution of the collective diffusion 
properties of a fluid immersed in an ideal porous medium using both protocols \cite{porous}.  
Moreover, there have been derived numerical criteria for the prediction of 
the liquid-glass transition for several systems and conditions \cite{rmf,laura4,rigo1,rigo2}, 
requiring as input only structural properties as the static structure factors, 
$S_{\alpha \beta}(k)$. 
Actually, it has been shown that, for some cases, this theory is as useful as MCT in the 
localization of dynamic arrest transitions. 
Since SCGLE theory has been applied succesfully in the description of the dynamics of 
colloidal fluids permeating ideal porous media as well as in the prediction of the 
liquid-glass transition in mixtures, it is obvious that the next step in this direction is 
the extension of this theory to predict the dynamic arrest transitions of QA and EM systems.
This is the aim of this work.

% \section{Preliminary concepts}

The relevant dynamic information of an equilibrium $\nu$-component
colloidal suspension is contained in the $\nu \times \nu$ matrix $F(k,t)$ whose elements are the {\it partial
intermediate scattering functions} $F_{\alpha \beta}(k,t)\equiv \left\langle n_{\alpha}({\bf k},t)
n_{\beta}(-{\bf k}^{\prime},0)\right\rangle$ where $n_{\alpha}({\bf k},t)\equiv \sum_{i=1}^{N_{\alpha}}
\exp[i{\bf k}\cdot {\bf r}_i(t)]/\sqrt{N_\alpha}$, with ${\bf r}_i(t)$ being the position of particle $i$ of
species $\alpha$ at time $t$. The initial value $F_{\alpha \beta}(k,0)$ is the partial static structure factor
$S_{\alpha \beta}(k)$ \cite{HANSEN, NAGELE}. 

% The SCGLE theory is summarized by a self-consistent system of
% equations \cite{rigo1,marco2} 

The multi-component version of the SCGLE theory consists of a set of
exact time-evolution equations that governs the relaxation of the partial intermediate
scattering functions $F_{\alpha \beta}(k,t)$ and their self contrapart
$F_{\alpha\beta}^{(s)}(k,t)\equiv \delta_{\alpha \beta}\left\langle \exp {[i{\bf k}\cdot \Delta{\bf
R^{(\alpha )}}(t)]} \right\rangle $, where $\Delta{\bf R}^{(\alpha )}(t)$ is the displacement of any of the
$N_{\alpha }$ particles of species ${\alpha}$ over a time $t$, and $\delta_{\alpha \beta}$ is Kronecker's delta
function. The Laplace transform (LT) $F(k,z)$ of the
matrix $F(k,t)$, can be written as \cite{marco1},

\begin{equation}
F(k,z)=\left\{z+(I+C(k,z))^{-1}k^{2}DS^{-1}\right\}^{-1}S,\label{fdz}
\end{equation}
where the elements of the matrix $D$ are given by $D_{\alpha \beta}
\equiv \delta_{\alpha \beta} D^0_{\alpha}$, with $D^0_{\alpha}$
being the diffusion coefficient of species $\alpha$ in the absence
of interactions. This is related with the solvent friction
coefficient on an isolated particle of species $\alpha$,
$\zeta^0_{\alpha}$, through the Einstein relation,
$D^0_{\alpha}\equiv k_{B}T/\zeta^0_{\alpha}$. The elements
$C_{\alpha \beta}(k,z)$ of the matrix $C(z)$ are the LT of the
so-called irreducible memory functions $C_{\alpha \beta}(k,t)$
\cite{laura4}. The corresponding result for the
``self'' component, $F^{(s)}(k,t)$. Defined as 
$F_{\alpha\beta}^{(s)}(k,t)\equiv \delta_{\alpha \beta}\left\langle \exp {[i{\bf k}\cdot \Delta{\bf
R^{(\alpha )}}(t)]} \right\rangle $, where $\Delta{\bf R}^{(\alpha )}(t)$ is the displacement of any of the
$N_{\alpha }$ particles of species ${\alpha}$ over a time $t$, and $\delta_{\alpha \beta}$ is Kronecker's delta,
in Laplace space is

\begin{equation}
F^{(s)}(k,z)=\left\{z+(I+C^{(s)}(k,z))^{-1}k^{2}D\right\}^{-1},\label{fsdz}
\end{equation}
where the matrix $C^{(s)}(k,z)$ is the corresponding irreducible
memory function which is related with $C(k,z)$ by

\begin{equation}
C(k,z)= C^{(s)}(k,z) = \lambda(k)\Delta\zeta^*(z), \label{vineyard}
\end{equation}

The matrix $\Delta\zeta^*(z)$ in is a
diagonal matrix whose $\alpha$-th diagonal element,
$\Delta\zeta_\alpha^*(z)$, is the Laplace Transform of the time-dependent 
friction function of particles of species $\alpha$. 
Such function reads \cite{rigo1}

\begin{equation}
\Delta \zeta ^{*} _{\alpha}(t) =\frac{D^0_{\alpha}}{3(2\pi)^3}\int
d^3k k^2 [F^{(s)}(k,t)]_{\alpha\alpha} [c \sqrt{n} F(k,t) S^{-1}
\sqrt{n} h]_{\alpha\alpha}, \label{dz}
\end{equation}

with the elements of the $k$-dependent matrices $h$ and $c$ being
the Fourier transforms $h_{\alpha\beta}(k)$ and $c_{\alpha\beta}(k)$
of the Ornstein-Zernike total and direct correlation functions,
respectively. Thus, $h$ and $c$ are related to $S$ by $S =
I+\sqrt{n}h\sqrt{n} = [I-\sqrt{n}c\sqrt{n}]^{-1}$, with  the matrix
$\sqrt{n}$ defined as $[\sqrt{n}]_{\alpha\beta} \equiv
\delta_{\alpha\beta}\sqrt{n_\alpha}$.

As illustrated in Ref. \cite{rigo1}, the solution of the SCGLE theory provides the time and
wave-vector dependence of the dynamic properties of the system contained in $F(k,t)$ and $F^{(s)}(k,t)$. 
It also provides equations for their long-time asymptotic
values, referred to as non-ergodicity parameters, which play the role of order parameters for the
ergodic--non-ergodic transitions. The most fundamental of these results \cite{rigo1} is the equation
for the asymptotic mean squared displacement $\gamma_\alpha \equiv \lim_{t \to \infty} <(\Delta
\textbf{R}^{(\alpha)})^2>$, which reads,

\begin{eqnarray}
\frac{1}{\gamma_{\alpha}}=\frac{1}{3(2\pi)^3}\int d^3k k^2 
\left\lbrace\lambda [ \lambda  +k^2\gamma ]
^{-1}\right\rbrace_{\alpha\alpha}\\ \nonumber
\times \left\lbrace c \sqrt{n} S \lambda[S \lambda +k^2\gamma ] ^{-1} \sqrt{n} h\right\rbrace
_{\alpha\alpha}\label{1}
\end{eqnarray}
where $S$ is the matrix of partial static structure factors, $h$ and
$c$ are the Ornstein-Zernike matrices of total and direct
correlation functions, respectively, related to $S$ by $S =
I+\sqrt{n}h\sqrt{n} = [I-\sqrt{n}c\sqrt{n}]^{-1}$, with  the matrix
$\sqrt{n}$ defined as $[\sqrt{n}]_{\alpha\beta} \equiv
\delta_{\alpha\beta}\sqrt{n_\alpha}$, and  $\lambda(k)$ is a
diagonal matrix given by $\lambda_{\alpha\beta} (k) = \delta_{\alpha
\beta} [1+(k/k^{(\alpha)}_c)^2]^{-1}$, where $k^{(\alpha)}_c$ is the
location of the first minimum following the main peak of $S_{\alpha
\alpha}(k)$.

To generalize the SCGLE theory to calculate the dynamic properties $F_{\alpha\beta}(k,t)$ 
of a colloidal fluid mixture of $\mu$ components immerzed in a porous matrix, we use the same
model employed in Ref. \cite{porous}. This consists of a $\nu$-component colloidal mixture where
a fraction of the particles, $\mu$ of the $\nu$ species, are diffusing in the random matrix of obstacles
formed by restant $\nu-\mu$ immobile (self-diffusion coefficients identically zero)
especies which play the roll of porous medium.
The derivation of corresponding equations to calculate the partial intermediate scattering functions
$F(k,t)$, $F^(s)(k,t)$ and the long time mean squared displacement
$\gamma_\alpha \equiv \lim_{t \to \infty} <(\Delta\textbf{R}^{(\alpha)})^2>$ of the
mobile species is inmediate. Simply we make the self-diffusion coefficients $D_{\alpha}^0=0$
of all species that represents the immobile particles in the system of Eqs. \ref{fdz},\ref{fsdz} and \ref{dz},
the result is a reduced set of $\mu$ equations for the $\mu$ mobile species, with the 
same functional form.

We will consider a binary hard sphere mixture of $N_1$ mobile particles of diameter 
$\sigma_1$ and $N_2$ obstacle particles of diameter $\sigma_2$. 
The only macroscopic control parameters are
the volume fractions $\phi_i=\pi n_i \sigma_i^3/6$, where $n_i=N_i/N$ and $N=N_1+N_2$, 
and the size asimmetry defined as $\delta=\sigma_1/\sigma_2$. 
The information about the mobility of the particles is contained in the self diffusion coefficients, 
$D^0_\alpha$, so that, making $D^0_2=0$ in Eq. \ref{dz} implies that particles of species 2 remain immobile.
On the other hand, the structure of the porous medium is contained in the matrix of static structure 
factors, $S$. 
In practice, the selection of the protocol to generate the porous matrix
determines the way on calculating the elements of matrix $S$. 
In this work, we used replica Orstein-Zernike equations 
with  given as the input to obtain the rest of the elements of the matrix $S$.
Thus, for $S_{22}(k)$ we employed the static structure factor of a monodisperse system in the case of QA mixtures,
and a static structure factor of a binary mixture in the case of EM systems.

%At this point the dynamic character of both species is well established. 
%However an important factor that determines many of the features of a porous medium is the
%process in which the matrix is preparated. In this work we considered both QA and EM protocols described above. 

\begin{figure}[ht]
\begin{center}
\includegraphics[scale=0.25]{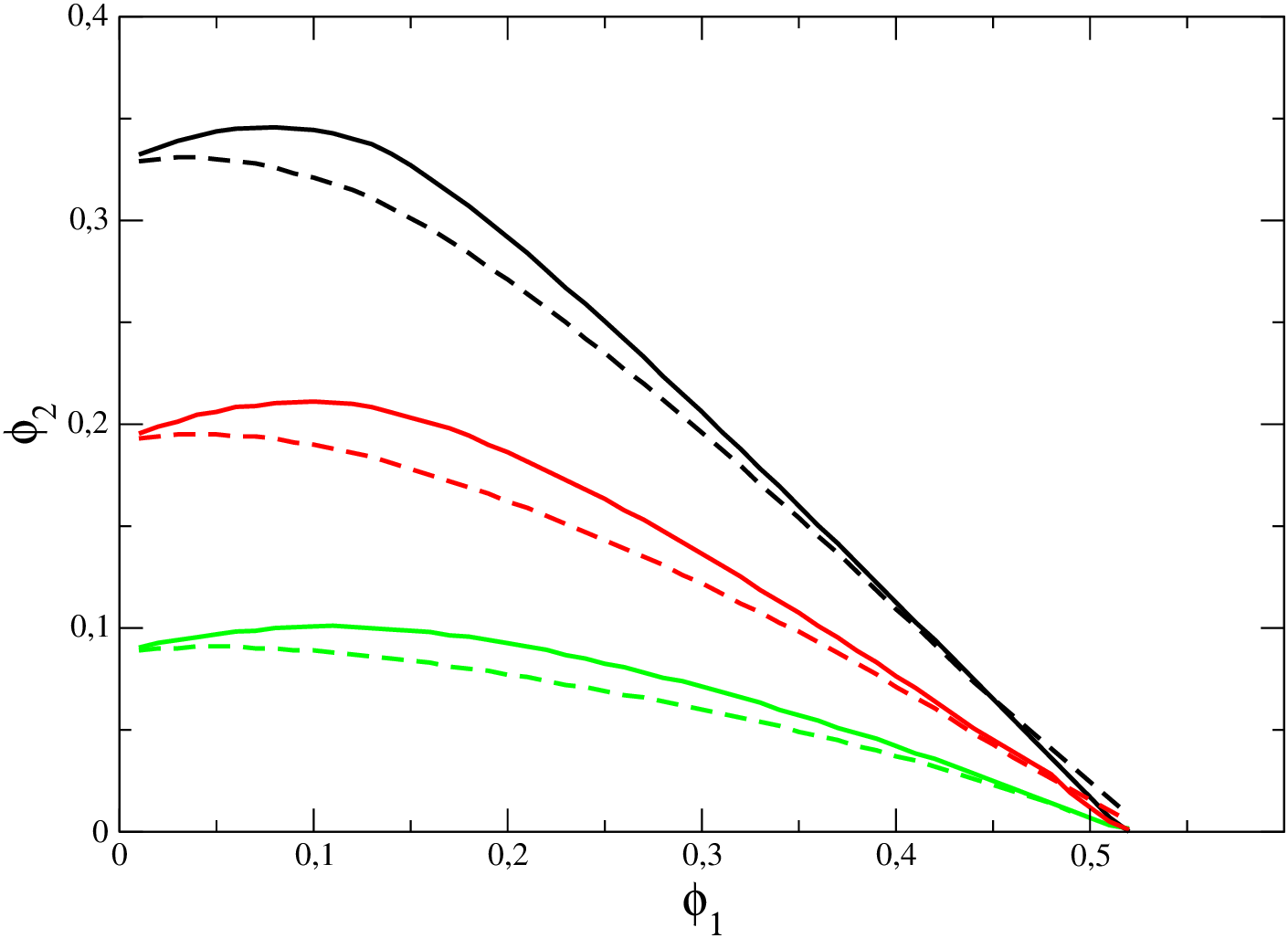}
\caption{Dynamic arrest phase diagrams of a monodisperse Hard Sphere fluid
immersed in a random porous medium prepared in two diferent models:
Quenched-Annealed (dashed lines) and Equilibrated-Mixture (continuous lines).
And three size asymetries ratios between fluid and matrix particles:
bottom $\delta=0.5$ (black), intermediate $\delta=1$ (red) and
top $\delta=2$ (green).(Color on line)
}
\label{fig.1}
\end{center}
\end{figure}

In Fig. \ref{fig.1} we compare the dynamic arrest diagrams obtained using Eq. \ref{1} 
with both protocols to explore the sensitivity of our theory to the morphology of the medium. 
Dashed lines correspond to Quenched-Annealed (QA) and
solid lines correspond to Equilibrated-Mixture (EM). Three size asymmetries
were studied: $\delta=0.5, 1$ and $2$; upper (black), middle (red) and
lower (green) curves respectively. Observe that EM protocol shows an evident reentrance while
for QA mixtures it is practically impossible to distinguish the existence of a reentrant region. 
We argue that this difference arises from the fact that thermalization process in EM systems, 
in which both species equilibrate together, leads to a self-generation of free-volume by
mobile particles. 
This is easy to understand considering the fact that an increase in the number of fluid 
particles for a fixed number of obstacles implies an increase in the total density 
which means less volume occupied by particle and so, in certain region of the transition
diagram, once the obstacle particles are quenched, the fluid has more free volume available to diffuse.
On the other hand, in QA systems, the volume occupied by the obstacles is the same independent of the 
concentration of fluid particles. 
So, an increase on the number of fluid particles would not increase their mobility, in principle, and 
for that reason there is not a reentrant region for QA systems.
These predictions are in agreement with computer simulation results reported in Refs. \cite{kim,kurzidim}.

\begin{figure}[ht]
\begin{center}
\includegraphics[scale=.25]{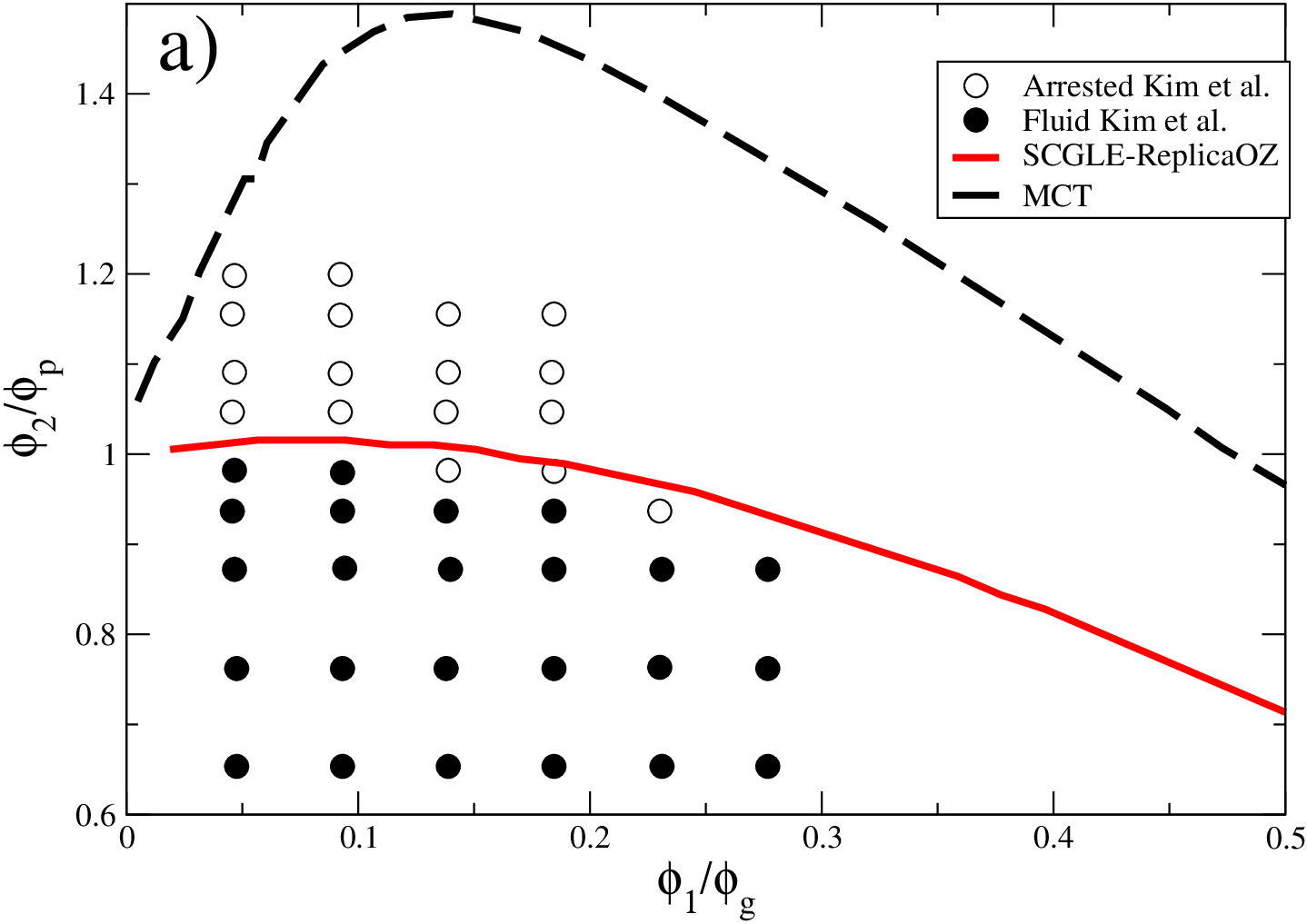}
\includegraphics[scale=0.25]{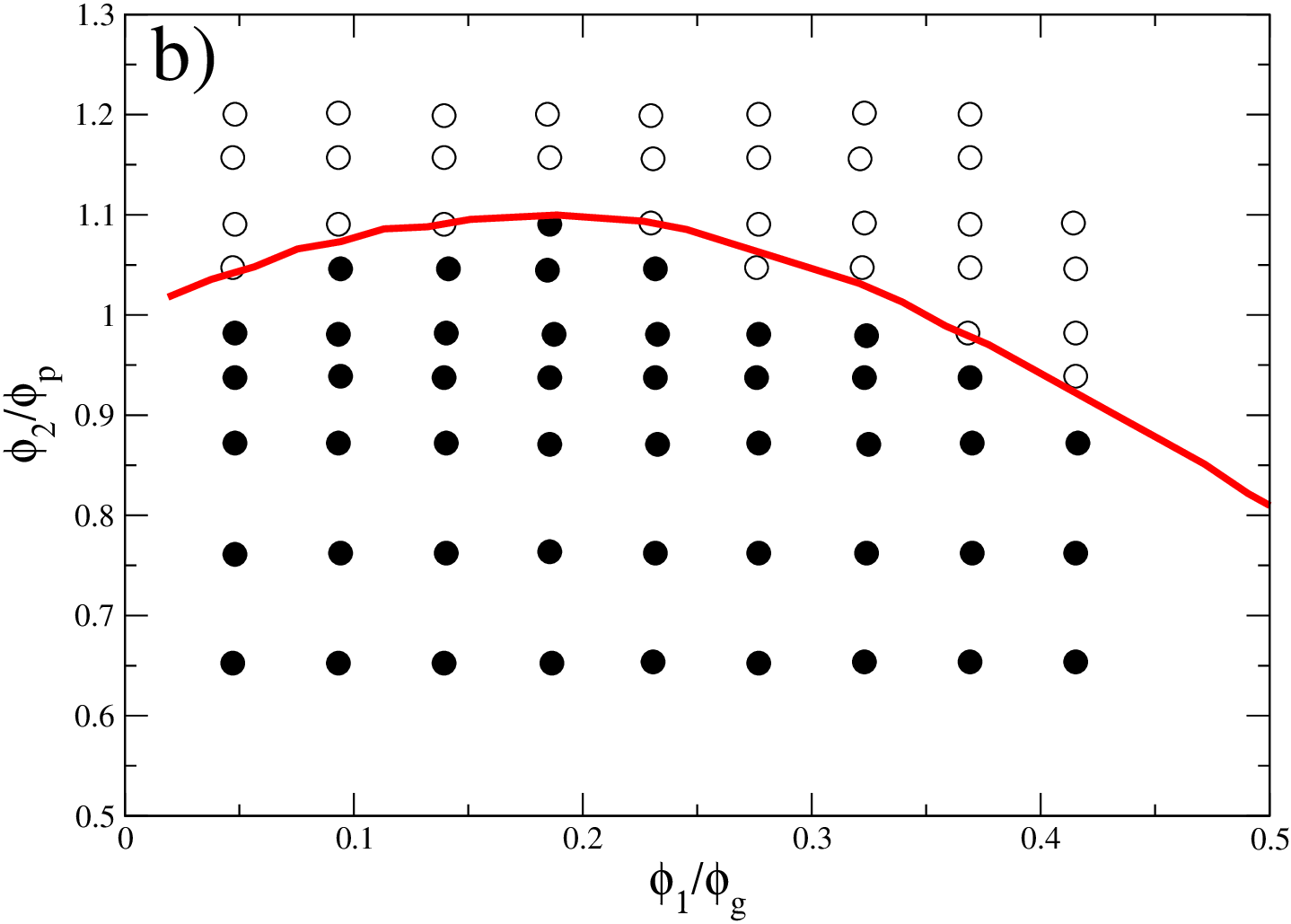}
\caption{Qualitative comparison between dynamic arrest diagram of hard sphere fluid
absorbed in a random porous medium calculated with SCGLE theory (continuous line)
and simulations of Kim et al (fill symbols fluid phase and empty symbols arrested phase),
for both QA (upper panel) and EM (lower panel) systems. 
RMCT prediction (dash line) is shown by reference.
}
\label{fig.2}
\end{center}
\end{figure}

Qualitative comparisons between SCGLE predictions and computer simulation results 
are shown in figure \ref{fig.2} where we plot the SCGLE transition line (continuous line) 
and the fluid (solid) and arrested (open) states determined by computer simulation 
experiments of Kim et. al. for QA mixtures (upper panel) and EM systems (lower panel). 
Vertical axis corresponds to the obstacle volume fraction scaled with the percolation 
volume fraction, $\phi_p$, and horizontal axis corresponds to the mobile particles 
volume fraction scaled with the monocomponent glass transition volume fraction, $\phi_p$.
For reference, we also plotted (dash line) the arrest transition line predicted 
by Krakoviack with RMCT for QA systems. 
It deviates considerably from simulation results and predicts a reentrancy that cannot be
appreciated in simulation experiments.
He explained this reentrancy in terms of the delocalization of mobile particles caused by 
occasional collisions with other fluid particles trapped in neighboor cages. 
We believe that this effect -if there exist- could not be responsible of the reentrant pocket, 
as it has been pointed out in Ref. \cite{kim} where the authors have shown that it is 
hard to identify the existence of a reentrant region in the arrest transition diagram obtained 
with molecular simulations. 

To understand the reentrant behavior of arrest transition diagrams with both protocols, 
one must carefully analyze the mechanisms 
that give rise to this phenomenon in both EM and QA systems.
First, Krakoviack proposed a mechanism to explain the notorious reentrancy in the diagram
predicted by RMCT with QA protocol \cite{krakoviack}, arguing that the dephasing of 
trajectories of the caged mobile particles by collision with other mobile particles
causes an increase in the percolation threshold. 
Although this mechanism could explain the existence of a reentrancy, says nothing
about the magnitude of this effect.
We think, in agreement with arguments of Kim et. al. \cite{kim}, that even when those collisions can eventually occur,  the structure
of the porous matrix is not modified and consequently the effects on the phase diagram should not be very noticeable. 
Thus, if this reentrant region does exist in the arrest diagram, the mesh size of the simulation
should be finer to show it clearly. 
This fact could be observed in Fig. \ref{fig.2} where the transition line predicted using SCGLE theory shows a barely noticeable reentrance. 
On the other hand, in the case of the EM protocol, the reentrancy is well understood and has already been explained in Ref. \cite{kim}.
The fact that both species (mobile and immobile) are equilibrated together before stopping the obstacle particles, 
implies that increasing the concentration of mobile particles, due to purely entropic effects, 
they generate their own space to continue moving. 
For that, unlike the case of QA, in EM systems 
the structure of the porous matrix is affected by this increase and so, the 
reentrance is hightly visible, as shown by MD results and theoretical (SCGLE) curve.

In summary, we have applied SCGLE to predict the dynamic arrest transition in a fluid 
permeating a porous matrix using Quenched-Annealed and Equilibrated Mixture protocols 
to generate the obstacle positions. 
As it has been shown in previous works, this 
theoretical scheme leads to an accurate description of the slow dynamics phenomena of 
fluids under different conditions. 
As can be seen in this work, SCGLE approach is capable to predict the arrest transition diagrams
in this kind of systems, independently of the protocol followed to generete obstacle positions.
The required input are the static structure factors of the mixture which could be calculated 
using ROZ equations with an apropriate closure relation, once the protocol is defined.
For EM systems, SCGLE arrest transition line is in excellent qualitative agreement with 
computer simulation results, showing a notorious reentrant pocket which 
is well understood and has already been discussed in this work and in Ref. [\cite{kim}].
On the other hand, for QA mixtures, we could compare our predictions with both 
computer simulations and another theoretical approach, Replica Mode Coupling Theory.
The qualitative agreement with computer simulation results is as good as in EM systems, 
but unlike RMCT predictions, the reentrant pocket is barely noticeable.
In this way, we think that SCGLE has provided us with a theoretical approach to 
predict dynamic arrest in mobile-immobile mixtures independently of the structure of 
the obstacle matrix in a unified way, just providing the appropriate static inputs.

This work was supported by Consejo Nacional de Ciencia y Tecnolog\'{\i}a through grants 
CB-2010-C01-156423 and Red Nacional de Nanociencias y Nanotecnolog\'{\i}a.

\end{document}